\documentclass[12pt,a4paper]{article}
\usepackage{epsfig}
\begin{document}

\begin{flushright}
{{\bf FIAN/TD/02/04}\\{\bf NETCOM-01/04}\\{\bf ITEP-TH-05/04}}
\end{flushright}

\medskip

\medskip

\begin{center}
{\bf ON NON-MARKOVIAN NATURE OF STOCK TRADING}\footnote{Presented at "Applications of Physics in Financial Analysis",
Warsaw, 13-15 November 2003}
\end{center}

\medskip

\begin{center}
{\bf Andrei Leonidov}
\end{center}

\medskip

\begin{center}

{\it (a) Theoretical Physics Department, P.N. Lebedev Physics Institute,\\
119991 Leninsky pr. 53, Moscow, Russia}

\medskip

{\it (b) Netcominvest Financial Investment Company,\\ 109017 Profsoyznaya 3, Moscow, Russia}\\

\medskip

{\it (c) Institute of Theoretical and Experimental Physics\\
117259 B. Cheremushkinskaya 25, Moscow, Russia}
\end{center}

\bigskip

\begin{center}
{\bf Abstract}
\end{center}

Using a relationship between the moments of the probability distribution of times between the two consecutive
trades (intertrade time distribution) and the moments of the distribution of a daily number of trades we show,
that the underlying point process generating times of the trades is an essentially non-markovian long-range memory
one. Further evidence for the long-range memory nature of this point process is provided by the powerlike correlation
between the intertrade time intervals. The data set includes all trades in EESR stock on the Moscow International
Currency Exchange in January 2003 - September 2003 and in Siemens, Commerzbank and Karstadt stocks traded on the
Xetra electronic stock exchange of Deutsche Boerse in October 2002 .

\newpage

One of the hot topics in the new field of econophysics \cite{MS00,BP01,S03} is working out a parsimonious
description of price dynamics. An ultimate challenge is to describe the whole price formation process starting from
intentions of market participants expressed in terms of buy or sell orders up to the bid - ask annihilation leading to
an appearance of the trade at given time, price and volume. Below we shall confine our consideration to the
"intermediate" level, where one looks only at the dynamics of accomplished trades. The material of this talk
is based on \cite{AL03}.

Among the most important characteristics of financial markets is their activity. On the trade-by-trade
basis one can think of two possible ways of characterizing it and thus setting the clock measuring
the operational time. The first possibility is to follow the volume traded \cite{C73}, the second is to analyze
the temporal pattern of trading operations. Below we shall follow the latter route.
The importance of effects related to the varying trading frequency is well recognized. One traditional topic of
great practical interest is a study of seasonality effects characterized by different timescales \cite{HFF01}.
On a more fundamental level, in \cite{PGAGS00} it was argued that long-range correlations in trading frequency
\cite{BLM00,PGAGS00} directly induce the observed long-range correlations of volatility.

Technically the influence of the temporal trading pattern on price dynamics can be taken into account by considering
the price generating process $P(t)$ as being subordinate \cite{F66} to the directing point process $T(t)$ that, in turn,
generates the times of the trades $t_1 < t_2 < \cdots\,$, so that $P(t)=P(T(t))$. Depending on the properties of $T(t)$,
probabilistic price dynamics for $P(t)$ can be described by differential (in time) or integral equations. More
specifically, one can distinguish the following generic types of $T(t)$:

\begin{itemize}

\item{{\bf Type 1.} The time at which the $n$-th point (trade) takes place is completely
independent of the times at which the previous $n-1$ points were generated. In this case one deals with the
Poisson distribution of the number of trades in any fixed time interval. This point process is fully characterized
by an {\it exponential} probability distribution for the intertrade time $\tau=t_n-t_{n-1}$,
$\psi(\tau)=\tau_0^{-1} {\rm exp} (-\tau/\tau_0)$. The evolution equations are then differential with respect to
time, and trading frequencies in non-overlapping time windows are independent.}

\medskip

\item{{\bf Type 2.} The time of the $n$-th trade is correlated with that of the previous
$n-1$-th trade, so that the the point process has a unit memory depth (and is thus, by definition, markovian).
The point process $T(t)$ is still fully characterized by some {\it non-exponential} intertrade time probability
distribution $\psi(\tau)$, but evolution equations are no longer differential, but integral ones, of
continuous time random walk (CTRW) \cite{MS84} type. The financial applications of CTRW were discussed in
\cite{SGM00,MRGS00,RSM02,MMG02,MMPW03,SGMMR03,RR03}. The correlation between the number of trades in
non-overlapping time windows can in principle be computed \cite{D50}, but the resulting expressions
are quite involved.}

\medskip

\item{{\bf Type 3.} The time of the n-th trade depends on $r>1$ times of the previous
trades $t_{n-1}, t_{n-2}, ..., t_{n-r}$. In this case the point process is a long-range memory non-markovian one with a
memory depth equal to $r$, and the corresponding evolution equations are complicated integral equations with
kernels depending on $r-1$ arguments. If all temporal scales are effectively involved, $r$ is infinite.}

\end{itemize}

\begin{figure}[h]
 \begin{center}
 \epsfig{file=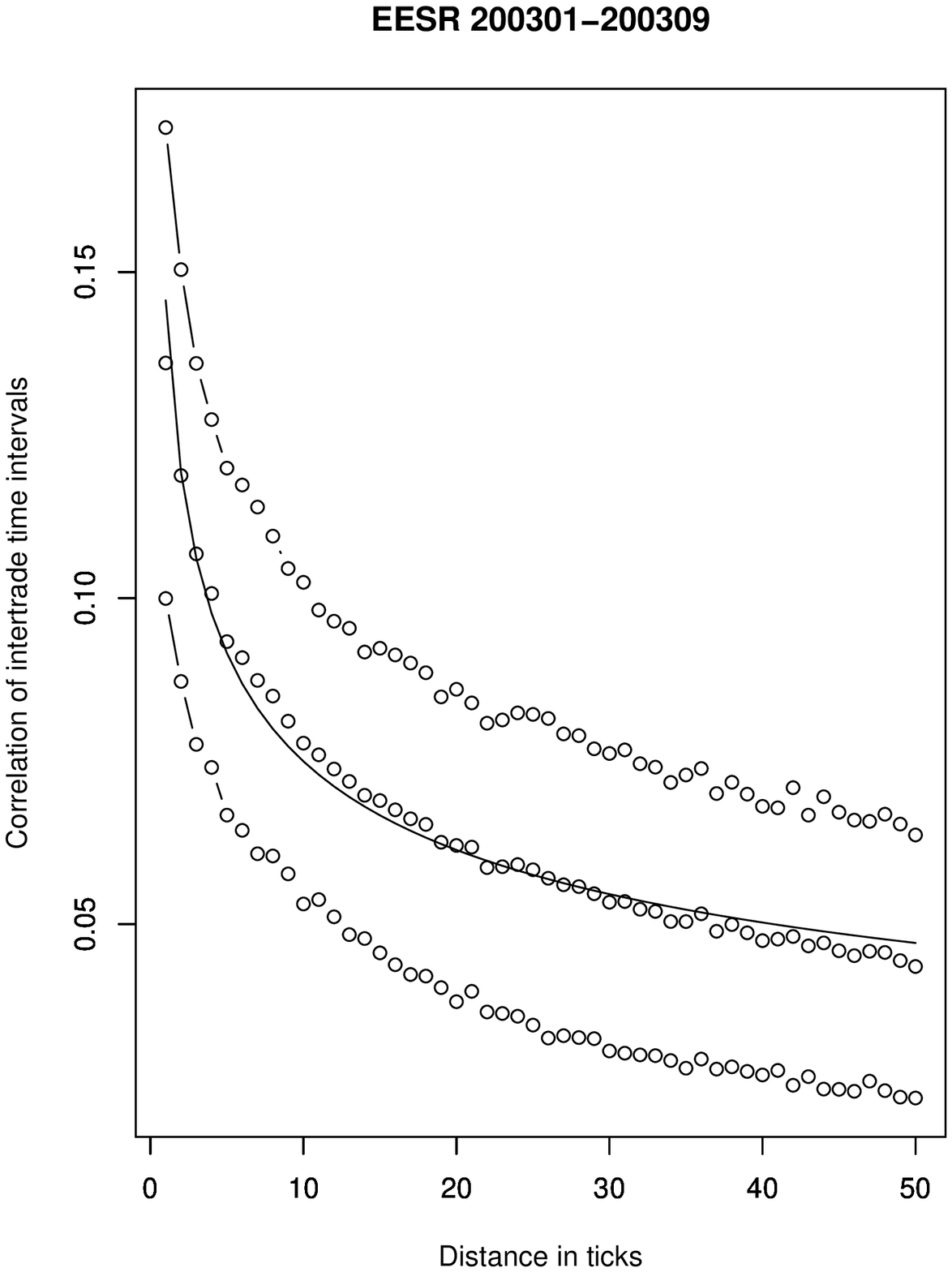,height=10cm,width=14cm}
 \end{center}
 \caption{Normalized correlation of intertrade time intervals in tick time. Solid line shows a powerlike
 fit of the form $0.15/d^{0.29}$. Upper and lower lines show the standard deviation of the measured correlation
 characterizing its variation on the day-to-day basis.}
 \label{tcor}
\end{figure}

Let us start our analysis of the character of the process generating the times of trades with studying the
normalized correlation of intertrade time intervals $\tau_i$ and $\tau_{i+d}$ separated by $d-1$ intermediate
intervals
\begin{equation}\label{ecor}
     C_{\tau}(d) \, = \, {\langle \tau_k \tau_{k+d} \rangle -\langle \tau \rangle ^2 \over \langle \tau^2 \rangle}
\end{equation}
The correlation is computed by first averaging over all intervals separated by a fixed number of intervals for a
given day and then averaging over all days. The correlation (\ref{ecor}) is shown, together with errorbars
characterizing its variation on the day-to-day basis, in Fig.~\ref{tcor}
From Fig.~\ref{tcor} one can conclude that the series of intertrade time intervals posesses long-memory
properties. The correlation function of intertrade time intervals $C_{\tau}(d)$ is slowly (powerlike) decaying,
$C_{\tau}(d) \simeq 0.15/d^{0.29}$, which is typical for long-range memory processes.

Another way of establishing the long-range memory nature of the point process generating the times of the trades
is to consider specific combinations of the moments of trade multiplicity distribution in
some given time interval $\Delta T$ and the moments of the intertrade time $\tau$ \cite{D50}. More specifically,
for the processes of Type 1 and Type 2 there exist, for some given large time interval $\Delta T$, a
relation between the first two moments of the distribution of the number of trades within this interval
$\langle N_{\Delta T} \rangle $ and $\langle N_{\Delta T}^2 \rangle $ and the first two moments of the intertrade
time distribution $\psi_1 \equiv \langle \tau \rangle $ and $\psi_2 \equiv \langle \tau^2 \rangle $ \cite{D50}:
\begin{equation}\label{ratio1}
 \langle N_{\Delta T}^2 \rangle  - \langle N_{\Delta T} \rangle ^2 = { \psi_2 - \psi_1^2 \over \psi_1^2 } \,
 {\Delta T \over \psi_1} \equiv
 { \psi_2 - \psi_1^2 \over \psi_1^2 } \, \langle N_{\Delta T} \rangle
\end{equation}
Corrections to (\ref{ratio1}) are of order $1/\Delta T$. Equivalently,
\begin{equation}\label{ratio2}
 \rho_N \equiv {\langle N_{\Delta T}^2 \rangle - \langle N_{\Delta T} \rangle ^2 \over \langle N_{\Delta T} \rangle }
 = { \psi_2 - \psi_1^2 \over \psi_1^2 } \equiv \rho_{\tau}
\end{equation}
so that if the underlying point process $T(t)$ is indeed of Type 1 or Type 2, one should have
$\rho_N/\rho_{\tau} = 1$. Let us note, that in the simplest poissonian case $\rho_N = \rho_{\tau} = 1$.

Let us now choose $\Delta T = 1$ day and consider all trades in  EESR in the period January 2003 - September 2003
and SIE, CBK and KAR in October 2002. Relevant information is summarized in Table 1:

\bigskip

\begin{center}
{\bf Table 1}

\medskip

\begin{tabular}{|c|c|c|c|c|c|}
\hline
 Und & $\langle N_{\Delta T} \rangle$ & $\langle \tau \rangle$ (seq)& $\rho_{\tau}$ & $\rho_N$ & $\rho_N/\rho_{\tau}$ \\
\hline
 EESR & 6823 & 4.5 & 10.3 & 492.8 & 109.4 \\
\hline
 SIE & 4364 & 9.1 & 3.8 & 124.8 & 32.8 \\
\hline
 CBK & 1856 & 21.3 & 3.1 & 274.4 & 88.5 \\
\hline
 KAR & 373 & 104.1 & 4.5 & 39.7 & 8.8 \\
\hline
\end{tabular}
\end{center}

From Table 1 it is clear, that poissonian Type 1 and CTRW Type 2 processes are excluded as
candidate point processes generating the times of the trades. This leaves us with the only
remaining possibility of Type 3 non-markovian long-range memory process. This conclusion is in
obvious agreement with the long-range correlation between the intertrade time intervals
for EESR shown in Fig.~(\ref{tcor}).

\begin{figure}[h]
 \begin{center}
 \epsfig{file=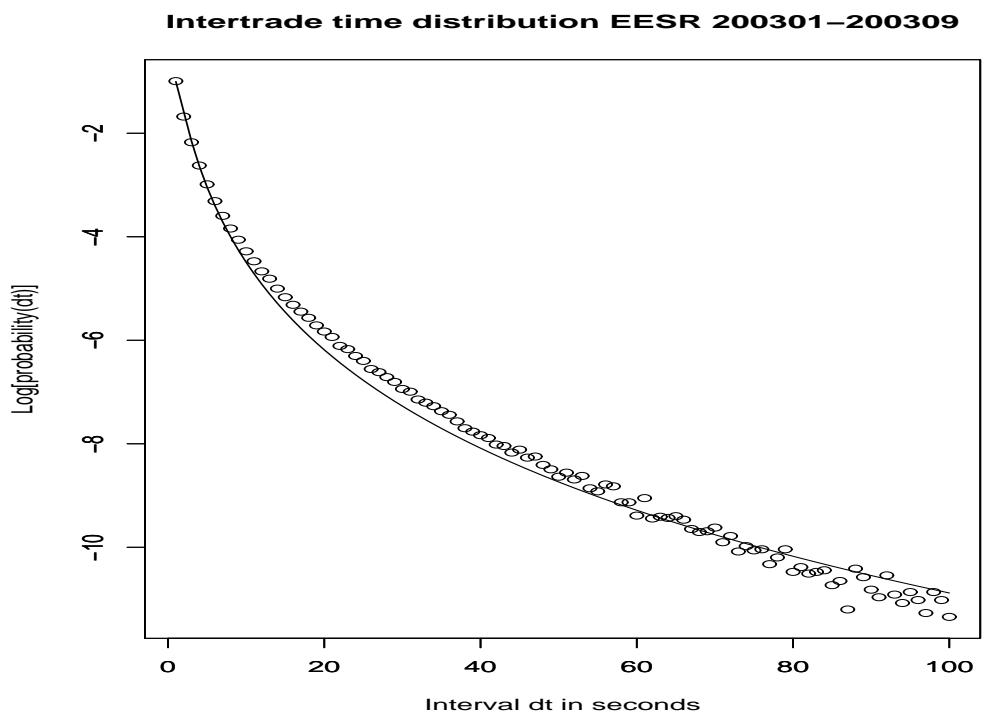,height=10cm,width=14cm}
 \end{center}
 \caption{Intertrade time distribution for EESR 200301-200309. Solid line shows a fit of the form
 $\exp \left \{  -(\ln \tau)^{1.5} -1 \right \}$}
 \label{dpdt}
\end{figure}

To provide some statistical background to this conclusion, let us consider a simulation of the
point process having exactly the same distribution over intertrade times as the EESR data shown
in Fig.~(\ref{dpdt}), but otherwise no additional correlations between the
intertrade time intervals \footnote{The distributions over intertrade time intervals were studied in
a number of papers \cite{RSM02,SGMMR03,RR03}}. The considered number of trade intervals $\Delta T$ was
equal to the number of trading days in the EESR data. This simulation effectively reconstructs a CTRW
process possessing the observed distribution over intertrade times.

The simulation results are shown in Table 2:

\begin{center}
{\bf Table 2}

\medskip

\begin{tabular}{|c|c|c|c|c|c|}
\hline
 Und & $\langle N_{\Delta T} \rangle$ & $\langle \tau \rangle$ (seq)& $\rho_{\tau}$ & $\rho_N$ & $\rho_N/\rho_{\tau}$ \\
\hline
 EESR (data)& 6823 & 4.5 & 10.3 & 492.8 & 109.4 \\
\hline
 EESR (sim.) & 6816 & 4.5 & 12.4 & 8.2 & 0.66 \\
\hline
\end{tabular}
\end{center}
We see, that although due to statistical limitations (finite size corrections, etc.)the simulation does not reproduce
the theoretical value of $\rho_N/\rho_{\tau}=1$, it is still close to it, while missing the experimental ratio by two
orders of magnitude.

\begin{center}
{\bf Conclusion}
\end{center}

We conclude that the point process generating the times of trades is of long-range memory non-markovian
nature.

\bigskip

This research was supported by RFBR grant 02-02-16779 and Scientific School Support grant 1936.2003.02.

\end{document}